%% file: scissors.tex
\documentclass[twocolumn,showpacs,preprintnumbers,prc]{revtex4-1}
\usepackage{aas_macros} 
\usepackage{graphicx}   
\usepackage{color}      
\usepackage{amsmath}    
\usepackage{amssymb}    
\usepackage{bm}         
\usepackage{float}      

\newcommand{\etal}{{\it et al.~}}                           
\newcommand{\nggn}{$(n,\gamma)\rightleftarrows(\gamma,n)$}  
\newcommand{\gsf}{$\gamma$SF}                               

\begin{document}

\title{Estimation of M1 scissors mode strength for deformed nuclei in the medium to heavy mass region by statistical Hauser-Feshbach model calculations}

\author{M. R. Mumpower}
\email{mumpower@lanl.gov}
\affiliation{Los Alamos National Laboratory, Los Alamos, NM 87545, USA}

\author{T. Kawano}
\affiliation{Los Alamos National Laboratory, Los Alamos, NM 87545, USA}

\author{J. L. Ullmann}
\affiliation{Los Alamos National Laboratory, Los Alamos, NM 87545, USA}

\author{M. Krti\v{c}ka}
\affiliation{Charles University, Prague, 180 00 Czech Republic}

\author{T. M. Sprouse}
\affiliation{University of Notre Dame, Notre Dame, IN 46556, USA}

\date{\today}
\begin{abstract}
Radiative neutron capture is an important nuclear reaction whose accurate description is needed for many applications ranging from nuclear technology to nuclear astrophysics. 
The description of such a process relies on the Hauser-Feshbach theory which requires the nuclear optical potential, level density and $\gamma$-strength function as model inputs. 
It has recently been suggested that the M1 scissors mode may explain discrepancies between theoretical calculations and evaluated data. 
We explore statistical model calculations with the strength of the M1 scissors mode estimated to be dependent on the nuclear deformation of the compound system. 
We show that the form of the M1 scissors mode improves the theoretical description of evaluated data and the match to experiment in both the fission product and actinide regions. 
Since the scissors mode occurs in the range of a few keV $\sim$ a few MeV, it may also impact the neutron capture cross sections of neutron-rich nuclei that participate in the rapid neutron capture process of nucleosynthesis. 
We comment on the possible impact to nucleosynthesis by evaluating neutron capture rates for neutron-rich nuclei with the M1 scissors mode active. 
\end{abstract}

\maketitle

\section{Introduction}
\label{sec:Introduction}
\input{introduction}

\section{Theory}
\label{sec:Theory}
\input{theory}

\section{Results and Discussions}
\label{sec:Result}
\input{result}

\section{Conclusion}
\label{sec:Conclusion}
\input{conclusion}

\section*{Acknowledgment}

We would like to thank A. Couture and P. Talou of Los Alamos National Laboratory and T. Kajino of National Astronomical Observatory of Japan for encouraging this work. 
This work was carried out under the auspices of the National Nuclear Security Administration of the U.S. Department of Energy at Los Alamos National Laboratory under Contract No. DE-AC52-06NA25396. 
M.K. acknowledges the support of the Czech Science Foundation under grant 13-07117S. 
TMS was supported by U.S. Department of Energy contract No. DE-SC0013039. 

\bibliographystyle{unsrt}
\bibliography{ref}

\end{document}

%% file: introduction.tex
The neutron radiative capture reaction on medium to heavy nuclei is a relatively simple process, compared to nuclear fission that involves a large number of degree-of-freedom to calculate \cite{Bohr1936}. 
Nevertheless, our capability to accurately calculate neutron capture cross sections is not yet well established, despite this process being one of the most important nuclear reactions for many applications from nuclear technology \cite{Huber2016} and nuclear based medicine \cite{Harling2003} to nuclear astrophysics \cite{Clayton1961}. 

In neutron radiative capture, an incoming neutron interacts with a target nucleus to form a compound nucleus (CN), which then decays by emitting $\gamma$-rays. 
The compound formation process is determined primarily by the optical potential, while the decay process is governed by two important nuclear properties, namely the nuclear level density (NLD) and the $\gamma$-ray (photon) strength function (\gsf). 
The latter two quantities are key ingredients in the statistical Hauser-Feshbach theory \cite{Hauser1952} for the neutron capture reaction, and accurate prediction of the radiative capture cross section relies on how well these nuclear properties are estimated. 
The level densities for stable nuclei are relatively well measured, as experimental data of average resonance spacing $D_0$ are available for many cases, and the level density can be extracted with an assumption for the spin and parity distributions. 

There are many available (almost exclusively phenomenological) \gsf \ models, see e.g. RIPL-3 \cite{RIPL3}, which can be used in calculation of the statistical decay. 
Both the phenomenological nature of the description and the variation between models contribute to the uncertainty in the prediction of capture cross sections \cite{Bartholomew1973} which is further compounded by the choice of implementation in Hauser-Feshbach codes \cite{Beard2014}. 
Among these models, one of the most widely used is the Generalized Lorentzian model of Kopecky \etal \cite{Kopecky1987}. 
A significant issue arises in the capture reaction calculation for deformed nuclei, where a standard Lorentzian form for the giant dipole resonance (GDR) for E1 transition often under-predicts measured capture cross section. 
Kopecky, Uhl, and Chrien \cite{Kopecky1993} proposed an enhanced generalized Lorentzian shape to expand the E1 GDR width to overcome this problem, although this representation includes an adjustable parameter. 
Watanabe \etal \cite{Watanabe2010} demonstrated a similar deficiency in calculating neutron capture cross sections for nuclei in the fission product (FP) region, and they reported that the available
Hauser-Feshbach codes tend to underestimate measured capture cross sections when re-normalization of $\langle\Gamma_\gamma\rangle$ to an experimentally or empirically estimated value is not performed. 

Recent attention to the $\gamma$-ray strength function has focused on the magnetic dipole excitation, M1. 
This mode was first observed in ground-state transitions in electron scattering experiments \cite{Bohle1984} and in photon scattering (also called nuclear resonance fluorescence) experiments \cite{Kneissl1996}. 
Studies performed by Ziegler \etal \cite{Ziegler1990} and Margraf \etal \cite{Margraf1993} laid the ground work for a correlation between M1 strength and nuclear deformation in the Sm and Nd isotopic chains respectively. 
The first observation of the M1 scissors mode in odd-A nuclei ($^{163}$Dy) was explored by Bauske \etal \cite{Bauske1993} and additional FP nuclei were studied by Pietralla \etal \cite{Pietralla1998}. 
Later, experimental data from neutron radiative capture reactions clearly indicated that the scissors mode is present in transitions between excited states at least up to transitions starting at neutron separation energy \cite{Krticka2004}. 
The interpretation of these findings was attributed to the M1 scissors mode which can be represented by a counter-rotational or out-of-phase oscillation of protons and neutrons in the nucleus. 
The amplitude of this collective oscillation is expected to be small at low excitation energies. 
For a comprehensive review of this subject see the article by Heyde, von Neumann-Cosel, and Richter \cite{Heyde2010}. 

As a result of these investigations, the M1 scissors mode needs to be considered in calculation of the radiative capture cross sections. 
In the past it was believed that M1 has a modest contribution to the calculated capture cross section, since the M1 spin-flip mode at around 7--10~MeV excitation is often under the tail of larger E1 GDR.  However, more recently, Ullmann \etal \cite{Ullmann2014} showed that the calculated neutron capture cross sections in the fast energy range are significantly enhanced by the addition of M1 strength. 
In this study of $^{238}$U, Ullmann \etal estimated the M1 scissors strength from the experimental capture cross section as well as the $\gamma$-ray multiplicity distributions measured with the DANCE (Detector for Advanced Neutron Capture Experiment) spectrometer at LANSCE (Los Alamos Neutron Science Center). 
A similar improvement when adding the M1 scissors mode was also reported by Guttormsen \etal for other nearby actinides \cite{Guttormsen2014}. 
The observation seen in the FP region \cite{Watanabe2010} is consistent with the finding for the actinide cases which suggests that the M1 scissors mode may impact capture cross sections of deformed or transitional nuclei throughout the chart of nuclides. 

In this work we investigate a correlation between the nuclear deformation and the scissors mode strength by analyzing the neutron capture data in the FP region, where a large variation of nuclear deformation is found. 
Since the fission channel is not involved, and charged particle emission is strongly suppressed by the Coulomb barrier, the nuclear reaction mechanism we have to deal with is relatively simple, and a standard Hauser-Feshbach theory with the coupled-channels framework \cite{Kawano2016} works well. 
From data gathered in the FP region, we posit a standard Lorentzian form for the M1 scissors mode and assume the center of the strength of the mode to be dependent on the nuclear deformation. 
This form of the M1 scissors mode yields a low energy increase to the \gsf \ in the range of a few keV to a few MeV. 
We find that this form of the M1 scissors seems to improve the theoretical description of both evaluated data and the match to experiment in the FP region. 
With the improved predictive capability in our statistical Hauser-Feshbach model, we expand our initial investigation outside the FP region to find regions where the M1 scissors mode may enhance capture cross sections and could be explored in future studies. 
We also report on the impact of the scissors mode on studies in astrophysics.

%% file: theory.tex
\subsection{Hauser-Feshbach theory for neutron radiative capture}

Here we consider neutron and $\gamma$-ray channels only. 
The Hauser-Feshbach formula with the width fluctuation correction for the neutron radiative capture process is written in the form
\begin{equation}
 \sigma_{\rm capt}(E_n) = 
    \frac{\pi}{k_n^2}
    \sum_{J\Pi} g_c \frac{T_n T_\gamma}{T_n + T_\gamma} W_{n\gamma} \ ,
 \label{eq:simgacapt}
\end{equation}
where $E_n$ is the incident neutron energy, $g_c$ is the spin statistical factor, $k_n$ is the wave-number of the incoming neutron, $T_n$ is the neutron transmission coefficient, $T_\gamma$ is the lumped $\gamma$-ray transmission coefficient, and $W_{n\gamma}$ is the width fluctuation correction factor. 
To calculate $W_{n\gamma}$, we use the model of Moldauer \cite{Moldauer1980} with the Gaussian Orthogonal Ensemble (GOE) parameterization \cite{Kawano2015}. 
The sum runs over the possible compound state spin $J$ and parity $\Pi$.

The capture cross section is related to the $\gamma$-ray strength function through the lumped $\gamma$-ray transmission as
\begin{equation}
  T_\gamma = \sum_{J'XL} \int_0^{E_0}
              2\pi E_\gamma^{2L+1}f_{XL}(E_\gamma) \rho(E_x,J') dE_x \ ,
\label{eq:Tgamma}
\end{equation}
where $E_0 = S_n + E_n$ is the total excitation energy of the compound nucleus, $E_x$ is the excitation energy of residual nucleus, $S_n$ is the neutron separation energy, $E_\gamma=E_0-E_x$ is the emitted $\gamma$-ray energy, $f_{XL}(E_\gamma)$ is the $\gamma$-ray strength function of multipolarity $L$ and type $X$ being E (electric) or M (magnetic), and $\rho(E_x,J)$ is the level density in the compound nucleus. 
The summation again runs over all allowed spin and parity combinations. 
When the final states of $\gamma$-decay are in discrete states, the integration in Eq.~(\ref{eq:Tgamma}) is replaced by a corresponding discrete sum. 
Note that in our actual calculation we do not use the lumped $\gamma$-ray transmission coefficient in Eq.~(\ref{eq:Tgamma}), but the continuum in a residual nucleus is discretized in order to calculate the width fluctuation correction properly \cite{Kawano1999b}. 

When an average $\gamma$-ray width $\langle \Gamma_\gamma \rangle$ is available from the experimental resonance parameters, the level density $\rho(E_x,J)$ and the lumped $\gamma$-ray transmission
$T_\gamma$ are connected by
\begin{eqnarray}
  T_\gamma &=& 2\pi \frac{\langle \Gamma_\gamma \rangle}{D_0} \ ,\\
  \label{eq:Gstrength}
  D_0 &=& 
  \left\{
    \begin{array}{ll}
      \frac{1}{\rho(S_n, 1/2)} & (I=0) \\
      \frac{1}{2} \left(
                  \frac{1}{\rho(S_n, I+1/2)} + \frac{1}{\rho(S_n, I-1/2)}
                  \right)  & (I\ne 0) \\
    \end{array}
  \right. \ ,
  \label{eq:d0}
\end{eqnarray}
where $I$ is the target nucleus spin, and $D_0$ is the average resonance spacing for $s$-wave neutrons. 
Here we omitted the trivial parity selection. 
This relation, however, is not fulfilled due to inconsistency between the employed $\gamma$-ray strength function and $\langle \Gamma_\gamma \rangle$. 
Under such circumstances, an empirical re-normalization is applied to the strength function. 

\subsection{$\gamma$-ray strength function for deformed nuclei}

The Hauser-Feshbach theory smooths out the detailed energy dependence of $f_{XL}(E_\gamma)$ by the integration of Eq.~(\ref{eq:Tgamma}). 
This integration also couples the NLD, thus making it difficult to extract an exact functional form of $f_{XL}(E_\gamma)$ from capture cross section data alone. 
Despite this, a rough estimate for the magnitude of strength function up to the neutron separation energy can still be obtained by comparing experimental neutron capture data. 

For the largest E1 giant dipole resonance (GDR), we adopt the generalized Lorentzian form of Kopecky and Uhl \cite{Kopecky1990}
\begin{eqnarray}
 f_{\rm E1}(E_\gamma) &=& 8.67 \times 10^{-8} \sigma_{\rm E1} \Gamma_{\rm E1} \nonumber \\
  &\times&
     \Large\{
     \frac{E_\gamma \Gamma_K(E_\gamma,T)}
          {(E_{\rm E1}^2 - E_\gamma^2)^2 + E_\gamma^2 \Gamma_K(E_\gamma,T)^2} \nonumber \\
  &+& 0.7 \frac{\Gamma_K(0,T)}{E_{\rm E1}^3}
     \Large\}
  \quad \ ,
  \label{eq:GLO}
\end{eqnarray}
where $E_\gamma$ is the energy of the $\gamma$-ray and $E_{\rm E1}$, $\sigma_{\rm E1}$, and $\Gamma_{\rm E1}$ are the GDR parameters. 
The units of the numerical constant are mb$^{-1}$MeV$^{-2}$, $\sigma$ in mb, and the units of width and energy are in MeV. 
This leads, in general, to the strength function in units of MeV$^{-(2L+1)}$. 
The temperature dependent width $\Gamma_K(E,T)$ is characterized by the level density parameter $a$ as
\begin{eqnarray}
  \Gamma_K(E_\gamma,T) &=& \left(E_\gamma^2+4\pi^2 T^2\right) \frac{\Gamma}{E^2} \ , \\
  T &=& \sqrt{\frac{S_n - E_\gamma}{a}} \ ,
  \label{eq:TdepGamma}
\end{eqnarray}
where $S_n$ is the neutron separation energy. 
The so-called enhanced generalized Lorentzian \cite{Kopecky1993}, which might be an alternative choice, is unsuitable for our purpose, since it already
includes an empirical enhancement in the deformed region. 
We employ the double-humped GDR parameters of Herman \etal \cite{Herman2013}. 

The expression in Eq.~(\ref{eq:TdepGamma}) is a little ambiguous when the $a$ parameter is energy dependent due to the shell-correction effect \cite{Ignatyuk1975, Ignatyuk1979}. 
The Hauser-Feshbach model code CoH$_3$ \cite{Kawano2010} employs the Gilbert-Cameron level density formula \cite{Gilbert1965}, which combines the constant temperature model $\rho_T$ at the low excitation energies and the Fermi gas model $\rho_G$ in the higher energy region. 
The shell correction mentioned above modifies the original Gilbert-Cameron level density in our calculation, and the model parameters have recently been updated \cite{Kawano2006}. 
The energy-dependent $a$ parameter is calculated as
\begin{eqnarray}
 a(U) &=& a^* \left\{
            1 + {{\delta W}\over{U}}\left( 1-e^{-\gamma U} \right)
         \right\} \ , \\
 \label{eq:levden}
  U &=& E_x - \Delta \ ,
 \label{eq:U}
\end{eqnarray}
where $\Delta$ is the pairing energy, $\delta W$ is the shell correction energy, $a^*$ is the asymptotic level density parameter, and $\gamma = 0.31A^{-1/3}$ is the damping factor. 
When the excitation energy $E_x$ is inside the constant temperature regime, the $a$ parameter is evaluated at the conjunction energy of $\rho_T$ and $\rho_G$.

Motivated by the observation of the M1 scissors mode in the actinide region \cite{Ullmann2014, Guttormsen2014}, we add a small Lorentzian to represent the scissors mode;
\begin{eqnarray}
  f_{\rm M1}(E_\gamma) &=& 8.67 \times 10^{-8} 
  \sigma_{\rm M1} \Gamma_{\rm M1} \nonumber \\
  &\times&
  {{E_{\rm M1} \Gamma_{\rm M1}}
   \over
    {(E_\gamma^2 - E_{\rm M1}^2)^2 + E_\gamma^2 \Gamma_{\rm M1}^2}} 
  \quad \ ,
  \label{eq:m1sf}
\end{eqnarray}
where $E_\gamma$ is the energy of the $\gamma$-ray and the other quantities are parameters of the scissors mode. 
For the location of M1 scissors, we assume a similar mass-dependence of $\propto A^{-1/3}$ to the other GDR's, as well as an assumption that the oscillation amplitude is proportional to the deformation parameter, $\beta_2$, in the compound nucleus. 
From our previous study on $^{238}$U \cite{Ullmann2014}, we have 
\begin{equation}
   E_{\rm M1} = 80 |\beta_2| A^{-1/3} \quad \mathrm{MeV} \ ,
   \label{eq:m1beta}
\end{equation}
which is similar to the theoretical prediction of $66 \delta A^{-1/3}$, where $\delta$ is the Nilsson deformation \cite{Bes1984}. 
While the calculated capture cross section is sensitive to this extra M1, it stays almost the same unless the product $\sigma_{\rm M1} \Gamma_{\rm M1}$ becomes very different. 
This product (or a Lorentzian area) is an adjustable parameter in our study. 
In addition to the E1 and M1 aforementioned, we include the standard Lorentzian profile for the M1 spin-flip mode and E2, with the systematic study on GDR in RIPL-3 \cite{RIPL3}.
The nuclear deformation parameter $\beta_2$ is taken from the Finite-Range Droplet Model (FRDM) \cite{Moller1995, Moller2012}. 

\subsection{Hauser-Feshbach calculation with M1 scissors mode}

In order to extract the M1 scissors mode strength in a wide mass range, we compare the Hauser-Feshbach calculation with the evaluated cross sections of FPs in ENDF/B-VII.1 \cite{ENDF7} and JENDL-4 \cite{JENDL4}, instead of directly comparing with the experimental data. 
The current conventional evaluation procedures do not include the M1 scissors mode. 
Instead, the calculated average $\gamma$-ray width in Eq. (\ref{eq:Gstrength}) is scaled to match the calculated capture cross section with the experimental data available at some neutron incident energies. 
This procedure gives a reasonable excitation function of neutron capture, while the strength of E1 GDR could be far from experimental photon absorption cross sections. 
Generally speaking the evaluated cross sections well represent experimental capture cross section data in the fast energy range whenever they are available, while the shape of the excitation function is often taken from the Hauser-Feshbach calculation. 
In other words, the evaluated data provide reasonable interpolation and extrapolation of experimental information by applying theoretical models.

\begin{figure}
  \begin{center}
  \resizebox{\columnwidth}{!}{\includegraphics{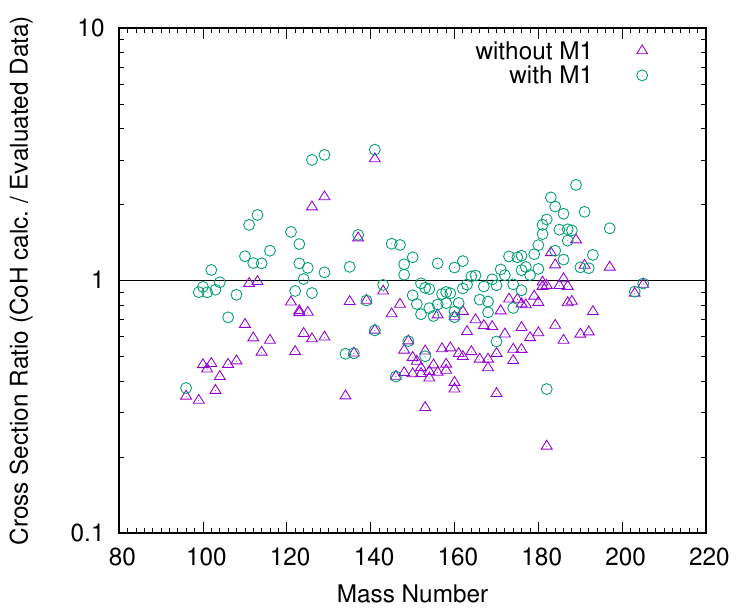}}
  \end{center}
  \caption{Ratios of the calculated capture cross sections at 200~keV to the selected evaluated cross sections in ENDF/B-VII.1 and JENDL-4. The triangles show the case when the Hauser-Feshbach calculations do not include the M1 scissors mode. The circles are with the scissors mode.}
  \label{fig:at200keV}
\end{figure}

We choose the neutron incident energy of 200~keV at which we compare our calculation with the evaluated values in the evaluated files. 
This energy could be higher than the resolved resonance region, and be lower than the region where the direct/semidirect capture process starts contributing to some extent. 
We select 106 nuclei in the FP region, for which evaluation of capture cross section is based on experimental data. 
If we employ a normal set of $\gamma$-ray strength functions, namely without the scissors M1, the calculated capture cross section at 200~keV often results in lower value than the evaluated cross section. 
This underestimation is shown in Fig.~\ref{fig:at200keV} labeled by ``without M1,'' as a ratio of calculated cross sections to the evaluated ones. 
We attribute this underestimation to the missing M1 strength coming from the nuclear deformation effect, although this assumption is somewhat crude. 
By adding an extra M1 strength given in Eq.~\ref{eq:m1sf}, we are able to estimate the missing strength required to reproduce the 200-keV data. 

The cross section calculation is performed with version $3.5.1$ of the CoH$_3$ statistical Hauser-Feshbach code \cite{Kawano2010}, which includes both the coupled-channels model for the neutron entrance channel and the Hauser-Feshbach model with width fluctuation correction for the statistical decay channels. 
The Engelbrecht-Weidenm\"{u}ller transformation \cite{Engelbrecht1973} is performed to take the direct inelastic scattering channels into account correctly in the width fluctuation calculation \cite{Kawano2016}. 
We adopt the global coupled-channels optical potential of Kunieda \etal \cite{Kunieda2007} for calculating the neutron transmission coefficients of deformed nuclei and otherwise use the Koning-Delaroche global optical potential \cite{Koning+03}. 

%% file: result.tex
\subsection{Estimation of M1 scissors strength}

\begin{figure}
  \begin{center}
  \resizebox{0.8\columnwidth}{!}{\includegraphics{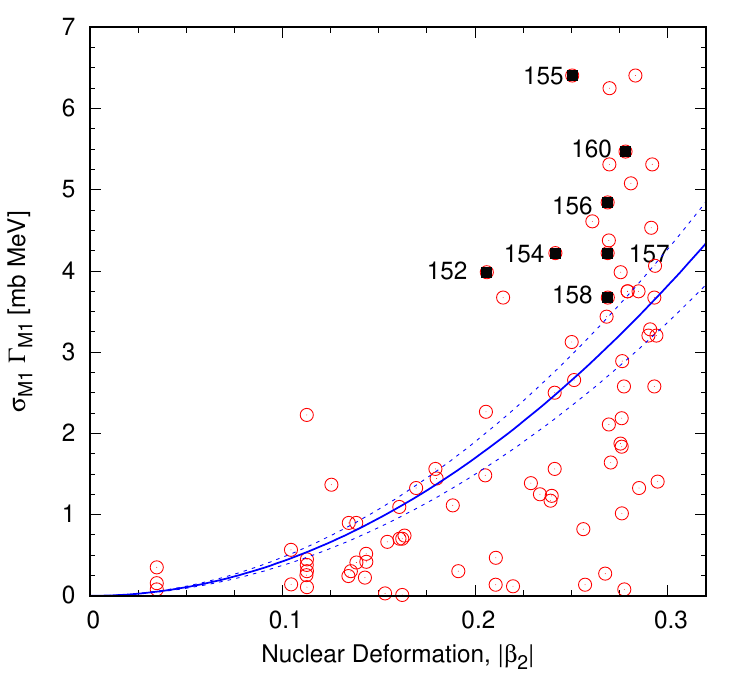}}
  \end{center}
  \caption{Additional M1 strength $\sigma_{\rm M1}\Gamma_{\rm M1}$ required to reproduce the evaluated capture cross section at 200 keV for selected nuclei in the fission product region. The quadratic curve is the least-squares fitting to the symbols, and the dashed curves are the 1-$\sigma$ band. The filled points are for the gadolinium ($Z=64$, Gd) isotopes.}
  \label{fig:m1search}
\end{figure}

The Lorentzian strength, $\sigma_{\rm M1}\Gamma_{\rm M1}$, is estimated by comparing the capture cross section at 200~keV, and the result is shown in Fig.~\ref{fig:m1search} as a function of the nuclear deformation $\beta_2$. 
Although the derived values are rather scattered, we can see some dependence on the deformation parameter. 
Assuming a quadratic form in $\beta_2$, which is reported by Ziegler \etal \cite{Ziegler1990}, Margraf \etal \cite{Margraf1993}, and Heyde \etal \cite{Heyde2010}, the least-squares fitting yields
\begin{equation}
  \sigma_{\rm M1} \Gamma_{\rm M1}
   = \left(42.4 \pm 5.0\right) \beta_2^2 \quad {\rm mb\ MeV} \ .
  \label{eq:m1systematic}
\end{equation}
Provided $\Gamma_{\rm M1}=1.5$~MeV we re-calculated the 200-keV capture cross sections with this relation and assuming, which are shown in Fig.~\ref{fig:at200keV} labeled by ``with M1.'' 
Overall the underestimated capture cross sections are reconciled, yet an overshooting is seen slightly above $A=180$.

\begin{figure}
  \begin{center}
  \resizebox{\columnwidth}{!}{\includegraphics{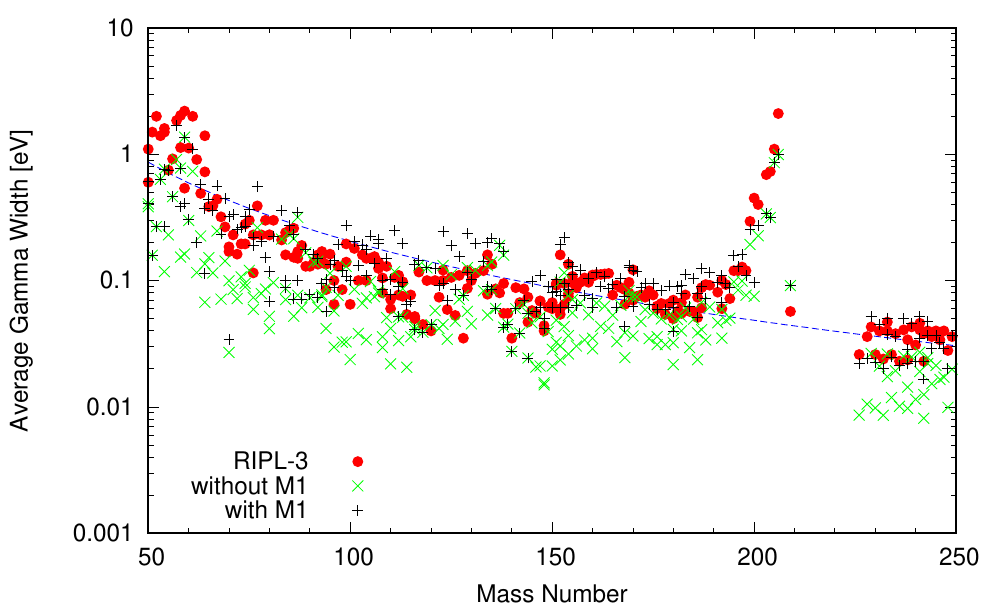}}
  \end{center}
  \caption{Comparison of the calculated average photon width $\langle\Gamma_\gamma\rangle$ with the compiled values in the RIPL-3 database. The filled circles are the evaluated values in RIPL-3, the $\times$ symbols are the calculated $\langle\Gamma_\gamma\rangle$ without the scissors mode, and the $+$ symbols are with the scissors mode. The dashed curve is a fit to the compiled $\langle\Gamma_\gamma\rangle$ values.}
  \label{fig:averagegamma}
\end{figure}

By summing up all the $\gamma$-ray strength functions including M1, the average $\gamma$-ray width $\langle\Gamma_\gamma\rangle$ is calculated by Eqs.~(\ref{eq:Tgamma}) and (\ref{eq:Gstrength}). 
We compare the calculated $\langle\Gamma_\gamma\rangle$ in the wider mass range with the resonance analysis values stored in RIPL-3 \cite{RIPL3}, which is shown in Fig.~\ref{fig:averagegamma}. 
Whereas the calculated $\langle\Gamma_\gamma\rangle$ values without the M1 strength are systematically lower than the resonance data in the mass $A=100 \sim 200$ region, inclusion of M1 improves this deficient
situation significantly, particularly in the mass $A=150\sim 200$ region where the nuclei tend to be deformed strongly. 
In the lower mass region, although our estimated M1 improves the agreement between the resonance data and the calculations, too much scatter in the data make our argument inconclusive. 
A possible impact from other photon strengths, such as the E1 pygmy dipole resonances, makes the situation more complicated in the low mass region.

The dashed curve in Fig.~\ref{fig:averagegamma} is represented by $\simeq 3000 A^{-2}$~eV which is a fit to the evaluated $\langle\Gamma_\gamma\rangle$ values. 
If we re-scale the calculated $T_\gamma$ to reproduce this simple relation, the calculated capture cross section won't be so unreasonable.  In fact, this technique is often adopted for estimating unknown capture cross sections. 
However, the re-scaling the whole $\gamma$-ray strength functions could cause unacceptable GDR parameters for E1. 
By adding M1, it is possible to avoid such artificial re-normalization of the $\gamma$-ray strength function.

\subsection{Comparison with experimental capture cross section data}
\subsubsection{Fission product region}

\begin{figure*}
  \begin{center}
   \centerline{\includegraphics[width=185mm]{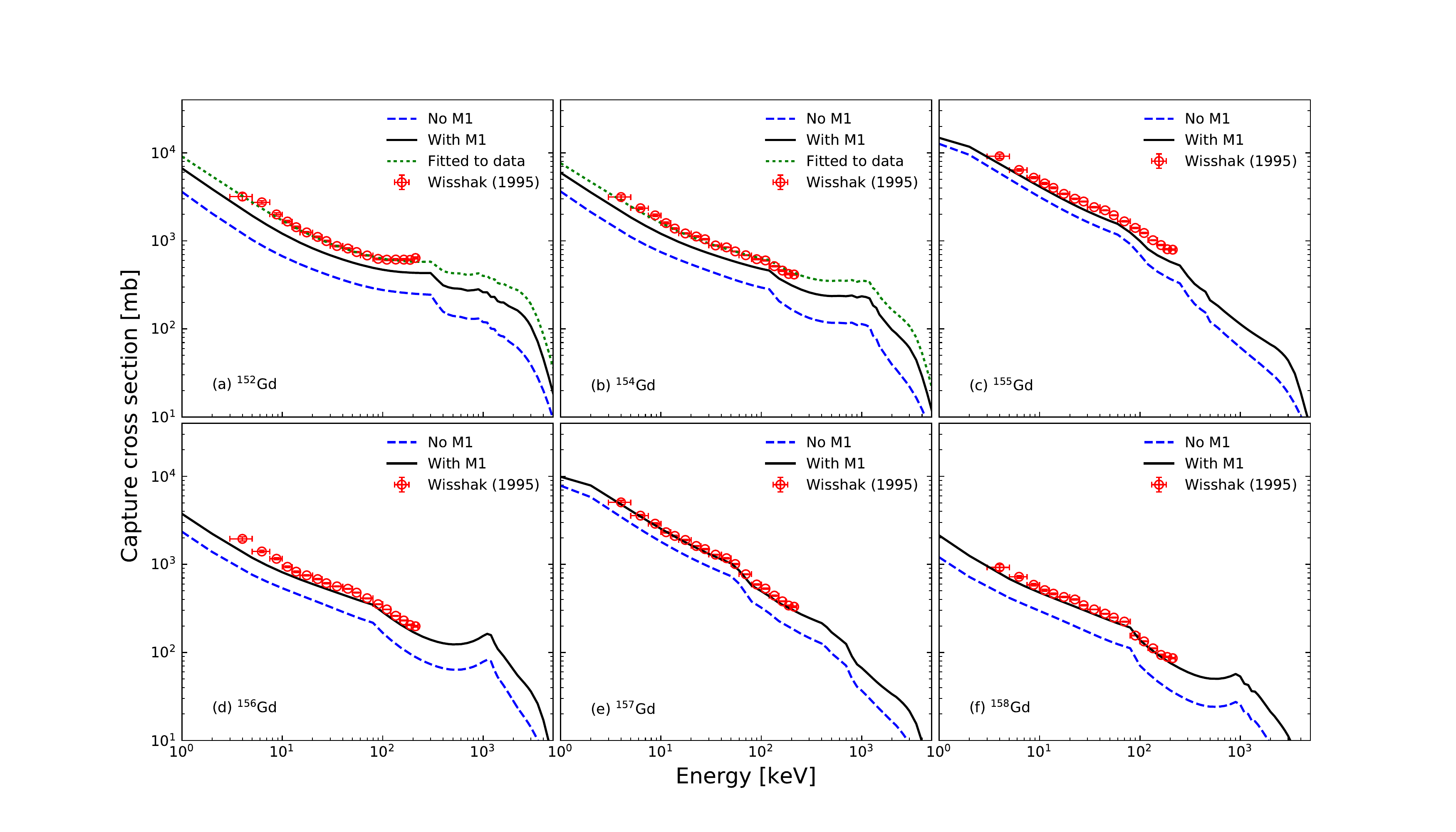}}
  \end{center}
  \caption{Comparison of the calculated capture cross sections with experimental data of Wisshak et al. \cite{Wisshak1995}. The dashed lines are calculated without M1 scissors, while the solid lines include M1. The dotted lines for $^{152}$Gd and $^{154}$Gd are the Hauser-Feshbach calculations fitted to the experimental data.}
  \label{fig:gdcapt}
\end{figure*}

The gadolinium ($Z=64$, Gd) isotopic chain has many highly deformed isotopes with measured capture cross sections making it an ideal candidate use for comparison between calculation and data in the FP region. 
The calculated capture cross sections for the Gd isotopes are compared with experimental data \cite{Wisshak1995} in Fig.~\ref{fig:gdcapt}. 
The agreement is significantly improved by including the the M1 strength of Eq.~(\ref{eq:m1systematic}) with assumed $\Gamma_{\rm M1}=1.5$~MeV. 
The lighter isotopes of $^{152}$Gd and $^{154}$Gd, for which deformation is smaller than heavier isotopes, seem to need more enhancement to reproduce the experimental data. 
As seen in Fig.~\ref{fig:m1search} by the filled data points, the Gd isotopes do not reveal clear $\beta_2$ dependency. 

By fitting the Hauser-Feshbach calculation to the experimental data of Wisshak \etal \cite{Wisshak1995}, we obtained $\sigma_{\rm M1} \Gamma_{\rm M1} = 4.1$ and 5.5~mb~MeV for $^{152}$Gd and $^{154}$Gd,
respectively. 
They are shown by the dotted curves in Fig.~\ref{fig:gdcapt}. 
These strengths are 2.2 times larger than the values given by Eq.~(\ref{eq:m1systematic}). 

There is no experimental data for $^{153}$Gd in the fast energy range. 
A recommended value of Maxwellian average cross section (MACS) in the KADoNiS database \cite{KADONIS} is 4550$\pm$ 700~mb at the temperature of 30~keV, while Eq.~(\ref{eq:m1systematic}) gives the MACS of 3322~mb. 
Considering the local deviation seen in $^{152}$Gd and $^{154}$Gd, we estimate $\sigma_{\rm M1}\Gamma_{\rm M1} = 4.3$~mb~MeV for $^{153}$Gd, and this yields the MACS of 3836~mb. 

\subsection{Comparison with DANCE multi-step cascade experiments}

\begin{figure}
  \begin{center}
  \resizebox{\columnwidth}{!}{\includegraphics{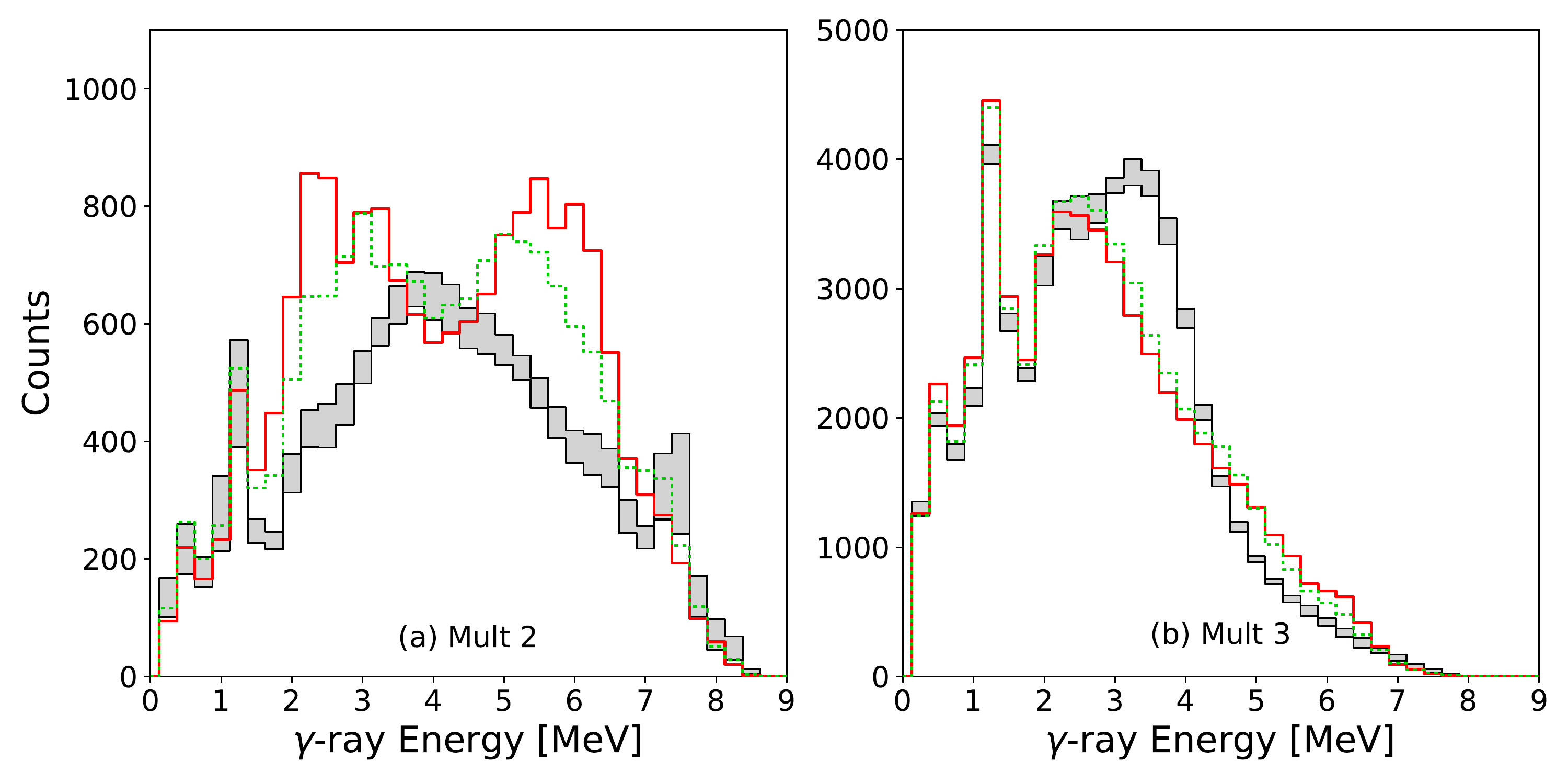}}
  \end{center}
  \caption{Calculated capture $\gamma$-ray spectra of the two step cascade (left panel) and the three step cascade (right panel) for the neutron-induced reaction on $^{155}$Gd, comparing the DANCE experimental data. The gray area is the simulated result, and the red and green histograms are the experimental $\gamma$-ray spectra from the two different resonances; the red is for the first $J=1^-$ resonance, while the green is for the 5-th $J=1^-$ resonance.}
  \label{fig:DANCE156}
\end{figure}

\begin{figure}
  \begin{center}
  \resizebox{\columnwidth}{!}{\includegraphics{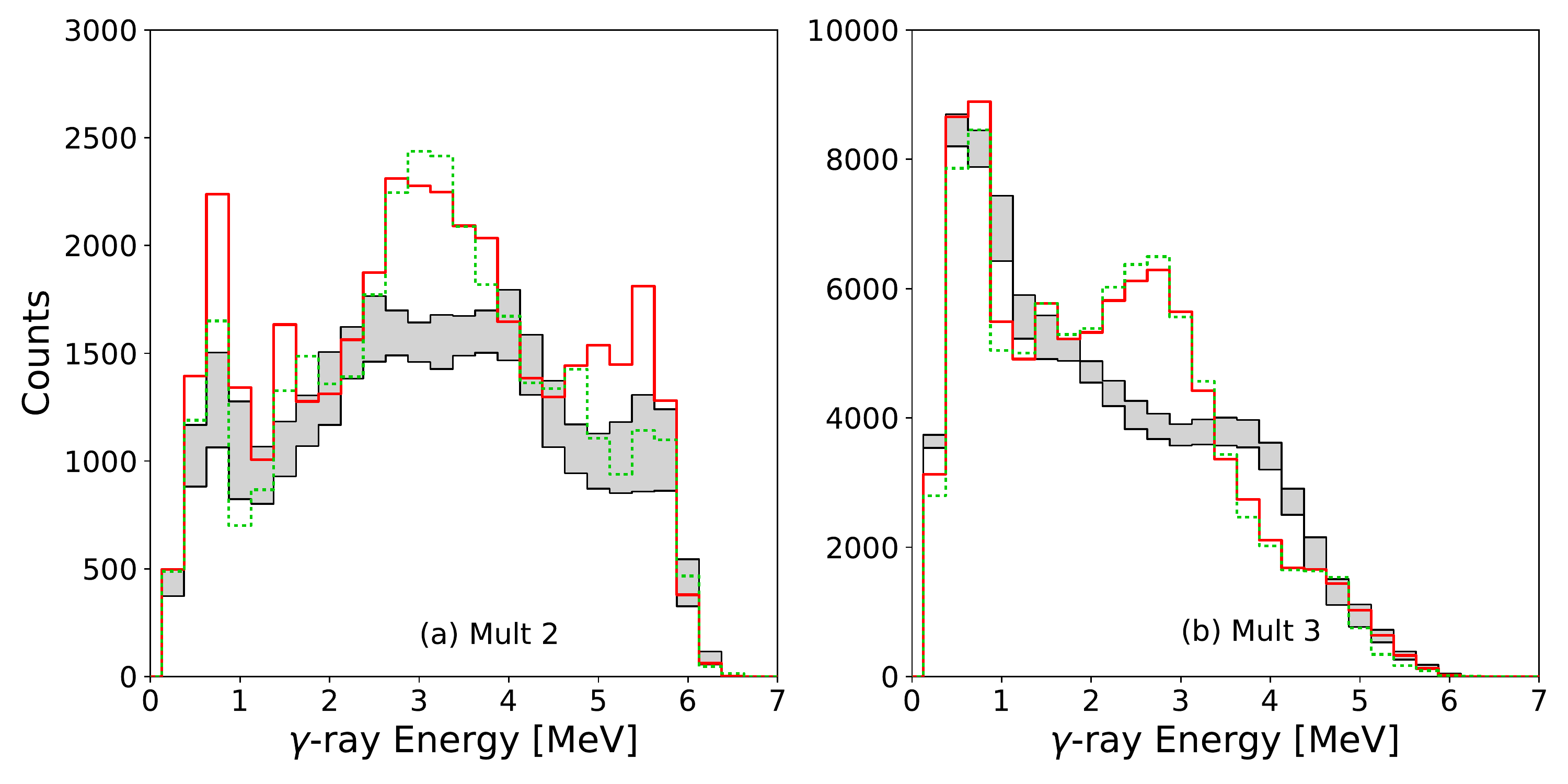}}
  \end{center}
  \caption{Calculated capture $\gamma$-ray spectra of the two step cascade (left panel) and the three step cascade (right panel) for the neutron-induced reaction on $^{156}$Gd, comparing the DANCE experimental data. The gray area is the simulated result, and the red and green histograms are the experimental $\gamma$-ray spectra from the two different resonances; the red is for the second $J=(1/2)^+$ resonance, while the green is for the 4-th $J=(1/2)^+$ resonance.}
  \label{fig:DANCE157}
\end{figure}

The behavior of the $\gamma$-ray strength in the MeV-region can be visible when the capture $\gamma$-ray spectrum is sorted by an individual multiplicity \cite{Ullmann2014}. 
We compare the estimated scissors mode strength with the DANCE experimental data of gadolinium isotopes \cite{Chyzh2011}. 
The multi-step cascade (MSC) simulation that includes the M1 strength was performed with the DICEBOX code. 
See Ref.~\cite{Chyzh2011} for the details of the calculation. 
The calculated two and three step cascade $\gamma$-ray spectra from $^{156}$Gd and $^{157}$Gd are compared with the DANCE experimental data in Figs.~\ref{fig:DANCE156} and \ref{fig:DANCE157}. 
The experimental data are shown by the red and green histograms, and the simulated MSC spectra are represented by the gray area. 
Although the simulated spectra in the 2--6~MeV range do not reproduce the experimental data, we must emphasize that the large enhancement in the spectra in that energy region cannot be obtained without the M1 scissors mode, and our estimation indeed moves the simulation toward the right direction. 

For the neutron-induced reaction on $^{155}$Gd, Eq.~(\ref{eq:m1beta}) suggests the M1 Lorentzian may locate at 3.7~MeV, which can be seen as a peak-position in the simulated spectra. 
The experimental data show that it could be in the 2--3~MeV region. 
We understand this shift is due to the uncertainty in the crude energy estimate in Eq.~(\ref{eq:m1beta}), and obviously a better estimation could improve the peak-position. 
The uncertainty in the M1 location also depends on $\beta_2$. 
When we assume $\beta_2=0.2$, a reduction in the deformation from the prediction of FRDM, the M1 Lorentzian will be shifted below 3~MeV, and the $M=2$ spectrum will be split into two-peaks as the experimental data. 

The calculated MSC spectra for the $^{156}$Gd reaction is in a similar situation. 
The estimated M1 position is higher by several hundred keV than the experimentally observed location. 
Because our aim is to estimate phenomenologically the global behavior of the M1 scissors mode without hefty computation that prevents large-scale nuclear data applications, we do not intend to fit the MSC spectrum to a particular experiment by adjusting the Lorentzian parameters in this study. 

\subsubsection{Extrapolation to actinide region}

\begin{figure}
  \begin{center}
  \resizebox{\columnwidth}{!}{\includegraphics{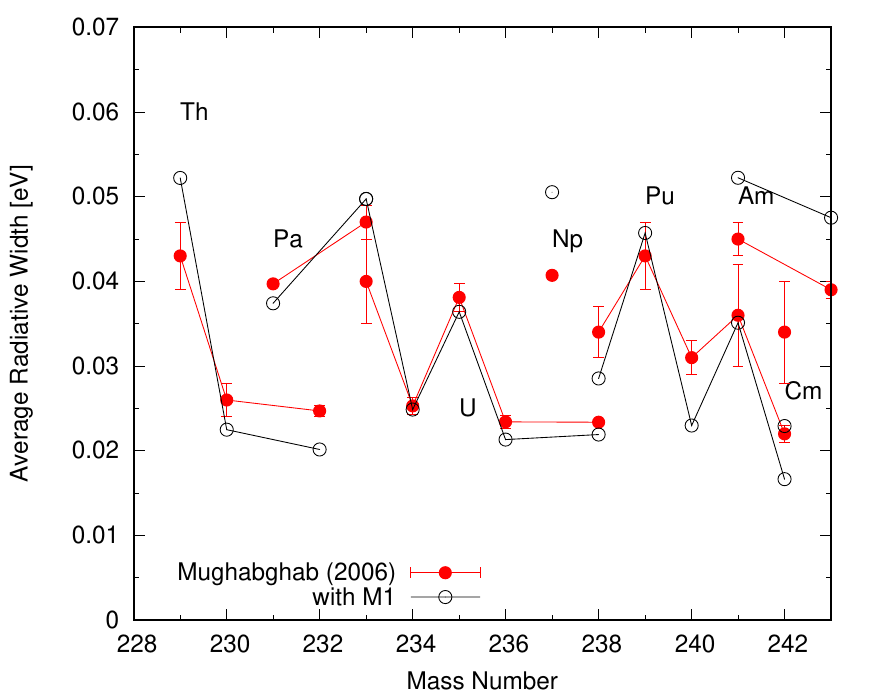}}
  \end{center}
  \caption{Comparison of the calculated average photon width $\langle\Gamma_\gamma\rangle$ with the compiled values in Ref. \cite{Mughabghab2006}. The filled circles are the evaluated values and the open circles are the calculated $\langle\Gamma_\gamma\rangle$ with the scissors mode.}
  \label{fig:averagegamma_actinide}
\end{figure}

Figure~\ref{fig:averagegamma_actinide} is a magnified plot of the $\langle\Gamma_\gamma\rangle$ comparison in the actinide region. 
The prediction of $\langle\Gamma_\gamma\rangle$ is also improved by adding M1. 
Here we compare the calculated $\langle\Gamma_\gamma\rangle$ with the compiled values in Ref. \cite{Mughabghab2006}. 
It is worth stressing that we derived the M1 scissors strength in the fission product region, and did not include any actinide data. 
Figure~\ref{fig:averagegamma_actinide} is a pure extrapolation to the heavier mass range. 
Although we overestimate the $\langle\Gamma_\gamma\rangle$ values by 15--25\% for $^{237}$Np and $^{241,243}$Am, our calculations are generally in good agreement with the evaluated values without any additional tweaks of model parameters. 
This also suggests the calculated capture cross sections with the M1 strength should reasonably reproduce the measured data in the fast energy region, when the fission channel is negligible. 
This is shown in Ref.~\cite{Ullmann2014}, although not exactly the same parameterization we proposed in this paper. 

\subsection{Application to neutron-rich nuclei}

We now explore the impact of including the M1 scissors mode to the neutron capture rates of neutron-rich nuclei and discuss the implications for the rapid neutron capture or $r$-process of nucleosynthesis. 

\subsubsection{M1 enhancement across the chart of nuclides}

\begin{figure*}
 \begin{center}
  \centerline{\includegraphics[width=190mm]{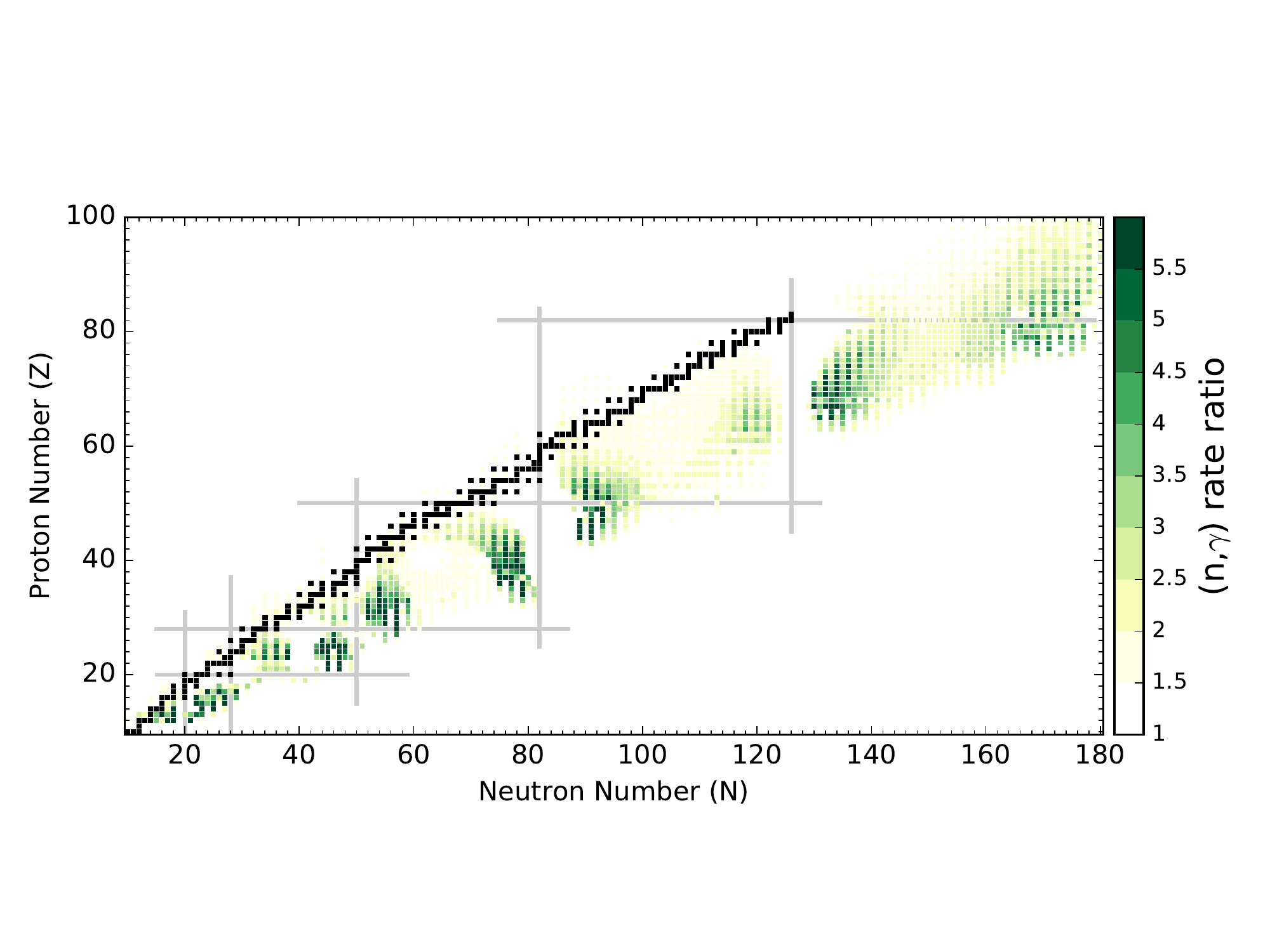}}
  \caption{\label{fig:ngratio} The M1 enhancement of neutron capture reaction rates evaluated at $T=1.0$ GK assuming the inclusion of the M1 scissors mode from Eq.~(\ref{eq:m1beta}). Each nucleus is colored based off the ratio of the new reaction rate (which includes the additional M1 strength) to the value of the old reaction rate (without M1 scissors mode). Darker shading represents a nucleus with a faster neutron capture rate when the M1 scissors mode is considered. Stable nuclei shown in solid black with closed shells denoted by gray solid lines.}
 \end{center}
\end{figure*}

Figure \ref{fig:ngratio} shows the ratio of neutron capture reaction rates with and without the M1 scissors mode. 
The ratio is calculated by taking the new reaction rate with M1 scissors mode and dividing by the old reaction rate without the additional M1 strength. 
For this figure, the neutron capture reaction rates are evaluated at $T=1.0$ GK, a rough estimate of the temperature the $r$ process may proceed through. 
With the M1 scissors mode active around $\sim$ 1 MeV, neutron capture rates of neutron-rich nuclei may increase up to a maximum factor of roughly $5$ at $T=1.0$ GK. 
We find that this enhancement from the M1 scissors mode has a larger impact for nuclei that are further from stability with the largest changes centralized in the transition regions just before or after closed shells, similar to previous predictions \cite{Schwengner+13}. 

At first glance, the distribution of increases to neutron capture rates in Fig.~\ref{fig:ngratio} may seem a bit peculiar as it does not follow in line with known maximums in $\beta_{2}$ deformation for
FRDM which occur near the mid-shells. 
The reason for this unintuitive spread relies on how neutron capture rates are calculated, recall Eq. (\ref{eq:simgacapt}). 
Neutron capture cross sections are proportional to $\frac{T_n T_{\gamma}}{T_n + T_{\gamma}}$, and to good approximation $T_n$ dominates the sum in the denominator, $T_n + T_{\gamma} \approx T_n$.
Thus, a relatively small change to the photon width can have a potentially large impact on the predicted neutron capture rates. 
Since this modification depends both on the strength and its placement in energy, which in our calculations are coupled to deformation, the distribution of impacted nuclei follows more closely the change to the average photon width, $\langle\Gamma_\gamma\rangle$, rather than following nuclei with largest deformation. 

\subsubsection{Application to nucleosynthesis calculations}

The astrophysical impact of additional low-lying $\gamma$-strength on the neutron capture rates of neutron-rich nuclei has been suggested by several groups \cite{Brown2014, Larsen2010}. 
In the case of Low Energy MAgnetic Radiation (LEMAR), the impact on the $r$ process was first considered in Ref.~\cite{Frauendorf2015}. 
In this study, systematic increases to neutron capture rates in a small region beyond the $N=82$ closed shell were shown to have a localized impact on the final isotopic abundances in the fission product region. 
Here we study the impact of the additional M1 scissors mode strength which is coupled to deformation as in Eq.~(\ref{eq:m1beta}). 
For each neutron-rich nucleus in the simulation, the neutron capture rate was calculated with and without the M1 scissors mode and a comparison was made between the baseline set of rates to the ones with the scissors mode. 

To study the nucleosynthesis we use a fully dynamical reaction network, Portable Routines for Integrated nucleoSynthesis Modeling (PRISM), which was recently developed at the University of Notre Dame \cite{Sprouse2017}. 
This network includes all relevant reaction channels from the JINA REACLIB database, support for changing any rate or property that goes into the network, as well as support for fission. 
The nuclear properties from FRDM are used in our network calculations as in Ref. \cite{Mumpower+15b}. 
By default the fission for these calculations uses an approximate asymmetric splitting schema \cite{Mumpower2017a} and is turned on for all astrophysical conditions. 
The region of nuclei which may fission in the $r$ process may only be reached in the case of extremely neutron-rich outflows. 

\begin{figure}
  \begin{center}
  \resizebox{\columnwidth}{!}{\includegraphics{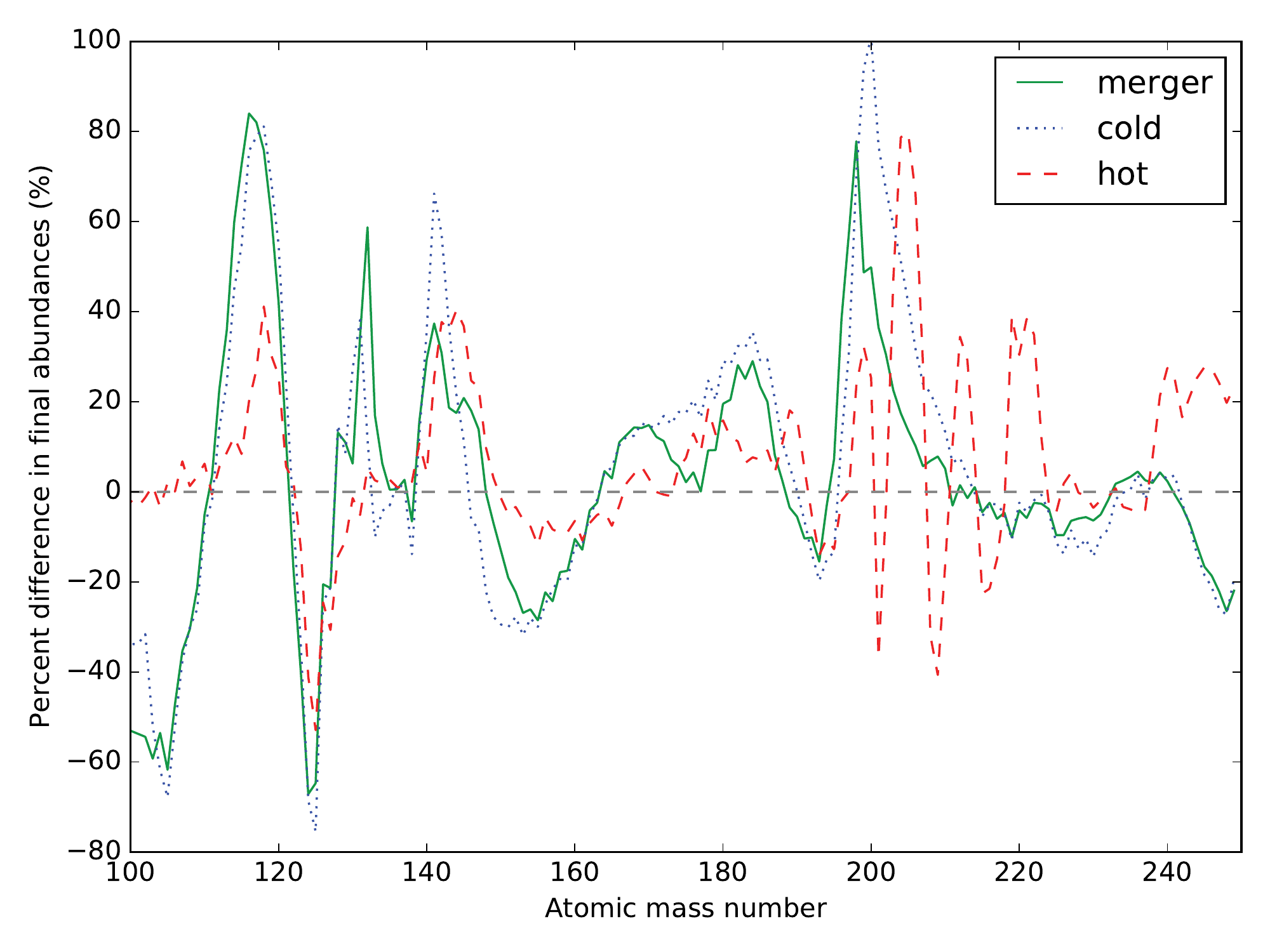}}
  \end{center}
  \caption{The change to final isotopic abundances when M1 enhancement is included in the neutron capture rates that go into $r$-process simulations. The behavior of the changes to the abundances is relatively similar, as denoted by the lines which represent simulations with different astrophysical trajectories. See text for details. }
  \label{fig:abpd}
\end{figure}

We consider three astrophysical conditions which include a low entropy hot $r$-process with a long duration \nggn \ equilibrium \cite{Mumpower+15b}, a cold $r$-process from a supernova scenario with some reheating \cite{Arcones+07} and a neutron star merger outflow from Ref. \cite{Goriely+11}. 
The impact of M1 enhancement for these three trajectories is shown in Fig.~\ref{fig:abpd}. 
We find the impact over a wide range of atomic mass units, with the cold and merger trajectories showing the most change from the baseline set of rates without the additional M1 scissors mode. 
We also note a slight boost in the production of the rare earth region just beyond $A\sim160$ is observed, which is often underproduced in $r$-process simulations relative to the larger peaks associated with closed neutron shells \cite{Mumpower+12c}. 

The constraint for an individual nucleus' neutron capture rate to alter the final abundances is that it is out of \nggn \ equilibrium and it is sufficiently populated during the decay back to stability. 
In addition, the population of the nucleus must also occur when neutron capture is still feasible, i.e, when neutron capture has not been limited by the amount of free neutrons or the astrophysical conditions. 
Further information on late-time neutron capture and its impact in the $r$-process can be found in Ref.~\cite{Mumpower+16r} and citations therein.

%% file: conclusion.tex
We studied the impact the M1 scissors mode has on neutron capture cross sections given a dependency on nuclear deformation. 
Assuming a simple Lorentzian form for the M1 scissors mode, the strength and position as a function of deformation was extracted by comparing the Hauser-Feshbach calculation with the evaluated nuclear data libraries that represent experimental data. 
The strength was found to be proportional to the square of the compound nucleus' deformation and the location in energy of this mode was assumed to be proportional to the absolute value of the compound nucleus' deformation, as in Ref. \cite{Ullmann2014} and previously predicted by Ref. \cite{Bes1984}.

We found the addition of the M1 scissors mode to the conventional generalized Lorentzian $\gamma$-ray strength function to provide an improvement in the prediction of neutron radiative capture cross section, especially for strongly deformed nuclei in the fission product region. 
This improvement is reinforced by considering the comparison of average $\gamma$-ray widths in the fission product region before and after the addition of the M1 scissors mode. 
The additional M1 dipole strength can also be applied to the nuclei in the actinide region, although, the lack of experimental data here currently limits any firm conclusions. 

We added the predicted M1 scissors mode to the calculation of neutron capture rates of nuclei throughout the chart of nuclides and explored the impact on $r$ process abundance yields. 
A speeding up, or `M1 enhancement', of neutron capture reaction rates was found for nuclei that lie in the transition regions before closed shells, but not at points of maximum deformation as predicted by the Finite-Range Droplet Model. 
The reason for this offset is because a neutron capture rate follows the change in average $\gamma$-width relative to the baseline Hauser-Feshbach calculation. 
We find the impact on the $r$-process abundance yields to be relatively small compared to the impact of other uncertain nuclear physics inputs, e.g. from nuclear masses \cite{Mumpower+16r}. 
Still, the impact is large enough that it should be taken into account when quoting the theoretical uncertainty of final abundances. 

Ongoing experimental campaigns and devoted theoretical efforts, particularly with large-scale shell model calculations, will continue to illuminate this phenomenon and guide us to understanding the complex, collective motion that occur in nuclei.